# Connectivity based Real-Time fMRI Neurofeedback Training in Youth with a History of Major Depressive Disorder

Xiaofu He[a,b,*], Diana Rodriguez Moreno[b], Zhenghua Hou[a], Keely Cheslack-Postava[a, b], Yanni Jiang[d], Tong Li[a], Ronit Kishon[a], Larry Amsel[a,b], George Musa[a,b], Zhishun Wang[a,b], Christina W. Hoven[a,b,c]

[a]Department of Psychiatry, College of Physicians & Surgeons, Columbia University, New York, NY 10032

[b]The New York State Psychiatric Institute, New York, NY 10032

[c]Department of Epidemiology, Columbia University, NY 10032

[d]Teachers College, Columbia University, NY 10032

[*]**Corresponding author**: Xiaofu He, E-mail: xh2170@cumc.columbia.edu

**Abstract**

***Background:*** Real-time functional magnetic resonance imaging neurofeedback (rtfMRI-nf) has proven to be a powerful technique to help subjects to gauge and enhance emotional control. Traditionally, rtfMRI-nf has focused on emotional regulation through self-regulation of amygdala. Although amygdala is a critical area in emotion processing, emotional control heavily depends on the connectivity between prefrontal and limbic areas. Furthermore, previous rtfMRI studies have observed that regulation of a target brain region is accompanied by connectivity changes beyond the target region. Therefore, the aim of present study is to investigate the use of connectivity between amygdala and prefrontal regions as the target of neurofeedback training in healthy individuals and subjects with a life-time history of major depressive disorder (MDD) performing an emotion regulation task.

***Method:*** Ten remitted MDD subjects and twelve healthy controls (HC) performed an emotion regulation task in 4 runs of rtfMRI-nf training followed by one transfer run without neurofeedback conducted in a single session. The functional connectivity between amygdala and prefrontal cortex was presented as a feedback bar concurrent with the emotion regulation task. Participants' emotional state was measured by the Positive and Negative Affect Schedule (PANAS) prior to and following the rtfMRI-nf. Psychological assessments were used to determine subjects' history of depression.

*Results:* Participants with a history of MDD showed a trend of decreasing functional connectivity across the four rtfMRI-nf runs, and there was a marginally significant interaction between the MDD history and number of training runs. The HC group showed a significant increase of frontal cortex activation between the second and third neurofeedback runs. Comparing PANAS scores before and after connectivity-based rtfMRI-nf, we observed a significant decrease in negative PANAS score in the whole group overall, and a significant decrease in positive PANAS score in the MDD group alone.

*Conclusions:* The present study suggests that healthy subjects could increase prefrontal cortex activity to regulate their emotions. However, the findings suggest that remitted MDD participants may require more sessions of the rtfMRI-nf training in order to learn to use the connectivity feedback to regulate their emotions and reduce negative feelings.

*Keywords:* Real-time fMRI Neurofeedback; Amygdala; Prefrontal Cortex; Functional Connectivity; Emotion Regulation; Major Depressive Disorder

## 1. Introduction

Real-time fMRI neurofeedback (rtfMRI-nf) has become a preferred technique in modulating cognitive behavior through direct observation of an individual's brain activity. Unlike traditional behavioral strategies, rtfMRI-nf provides subjects with immediate feedback and display of their current brain activation while performing a task, constituting a powerful neurofeedback signal of people's behavioral performance. Currently, most rtfMRI-nf studies have focused on the self-regulation of activity in one single brain region (Linden et al., 2012; Mehler et al., 2018; Ruiz et al., 2014; Young et al., 2017; Young et al., 2014; Zhu et al., 2019; Zotev et al., 2011; Zweerings et al., 2020), and the targeted region of interest (ROI) varies with task domain (Caria et al., 2007; Johnston et al., 2010; Linden et al., 2012; Mehler et al., 2018; Zotev et al., 2011). Emotion regulation, an essential skill for mental well-being and typically impaired in many psychological disorders (Gross, 2002; Campbell-Sills and Barlow, 2007), has been increasingly researched as one of the primary goals of rtfMRI-nf. Emerging evidence has confirmed rtfMRI-nf as a powerful technique in guiding individuals to gauge and enhance their emotional control (Herwig et al., 2019; Linden et al., 2012; Sarkheil et al., 2015; Zotev et al., 2013; Zweerings et al., 2020). Likewise, the majority of rtfMRI-nf studies aimed at emotional regulation deal with one brain area, typically the amygdala (Paret et al., 2014; Posse et al., 2003; Sarkheil et al., 2015; Yuan et al., 2014; Zotev et al., 2011; Zotev et al., 2013; Zweerings et al., 2020). Although the amygdala is a critical hub of emotion processing, it has been shown that normal emotion regulation requires the involvement of connected regions for the detection of emotionally salient stimuli, the generation of an emotional response, and a control system responsible for the voluntary or automatic regulation of those emotional responses (Banks et al., 2007; Lee et al., 2008; Morawetz et al., 2020). This emotional circuit comprises areas in the frontal regions, including the anterior cingulate cortex (ACC), dorsolateral prefrontal cortex (dlPFC), ventral and medial PFC (vmPFC), and orbitofrontal gyrus (Banks et al., 2007; Hariri et al., 2003; Morawetz et al., 2017; Ochsner et al., 2004; Stein et al., 2007). It has been proposed that the connectivity within the circuitry of prefrontal regions to the amygdala is more important than the activity of a single region during the processing of negative stimuli (Liddell et al., 2005).

Functional connectivity refers to the organization, inter-relationship, and performance of different brain regions (Rogers et al., 2007). Off-line post hoc analysis of the fMRI data in rtfMRI-nf has shown that self-regulation of a single ROI is accompanied by modulations in the connectivity of whole-brain networks (Hamilton et al., 2011; Lee et al., 2012; Lee et al., 2011; Rota et al., 2011; Ruiz et al., 2013; Scharnowski et al., 2014; Veit et al., 2012; Zilverstand et al., 2014; Zotev et al., 2011). Increasing evidence indicates that psychiatric disorders and their cognitive and emotional symptoms are linked to abnormal functional connectivity patterns (Bullmore, 2012; He et al., 2019; Ho et al., 2015; Kong et al., 2013; Tak et al., 2021; Zhang et al., 2020) rather than the dysfunction of a single brain region. Hence, there has been an increasing interest in using functional connectivity as a neurofeedback signal during real-time fMRI: Relevant studies have reported successful connectivity changes in motor or emotion networks among healthy subjects (Megumi et al., 2015; Tsuchiyagaito et al., 2021; Zhao et al., 2019; Zilverstand et al., 2014) as well as individuals with autism (Gotts et al., 2019; Ramot et al., 2017) after each regulation trial of neurofeedback learning. Concerning emotion regulation, changes in amygdala and connectivity with prefrontal regions following rtfMRI-nf training have been observed in post-analysis of healthy adult subjects (Herwig et al., 2019; Zotev et al., 2013), adolescents (Zich et al., 2018), and patients with Major Depressive Disorder (MDD; Young et al., 2018). More recently, in a proof-of-concept study, participants with subclinical depression were trained by functional connectivity neurofeedback between dlPFC and posterior cingulate cortex (PCC); promising outcomes including decreased depressive and brooding symptoms were observed (Taylor et al., 2022). Nevertheless, research into emotion regulation using connectivity-based rtfMRI-nf in psychological disorders such as MDD is still preliminary and limited.

MDD, with a lifetime prevalence rate of 16.2% in the United States (Kessler et al., 2003), is characterized by persistent deep sadness, reduced energy, loss of interest, disturbed sleep, vegetative nervous system dysregulation, cognitive dysfunction, and even a high suicidal tendency (American Psychiatric Association, 2013). Depression studies have revealed abnormalities in functional connectivity (task-related and resting-state) as well as structural connectivity in multiple brain networks (Anand et al., 2009; Connolly et al., 2013; Heller, 2016; Ho et al., 2015; Klauser et al., 2015; Sheline et al., 2010; Tak et al., 2021). Although many MDD patients could reach considerable improvement in depressive symptoms through effective therapy, residual symptoms can persist even after intensive treatments, leading to an increased risk of relapse shortly after remission (Paykel, 2022). Remitted MDD refers to a not fully recovered MDD state, during which individuals are still at fairly high risk of adverse outcomes, including future relapse, morbidity, and even mortality (Liu et al., 2015). The cumulative proportion of recurrence among recovered subjects could reach 85% (Mueller et al., 1999) and the relapse risk was evaluated to be around 60% when remitted patients suffered from two or more episodes (Bockting et al., 2015). Meta-analysis has indicated that individuals with remitted MDD, similar to current MDD patients, could suffer from maladaptive use of emotion regulation strategies (Visted et al., 2018). Aberrant resting-state connectivity patterns in multiple brain regions, such as the affective network (AN), the default mode network (DMN), and the cognitive control network (CCN), have been revealed among remitted depression patients through a series of studies (Stange et al., 2017; Zamoscik et al., 2014; also see Li et al., 2018 for review). Moreover, participants with remitted depression were uncovered to exhibit reversed and abnormal frontotemporal connections when viewing emotional faces; the connections between the orbitofrontal cortex and amygdala/fusiform gyrus were actively involved when remitted depression patients were viewing happy faces and when control participants were viewing sad faces (Goulden et al., 2012). Significantly increased caudate-amygdala and caudate-hippocampus functional connectivity was also detected in remitted MDD patients but not in healthy controls when confronted with negative stimuli (Admon et al., 2015).

Despite the significance of functional connectivity for emotion regulation in MDD and remitted MDD, to the best of our knowledge, no studies have directly addressed the self-regulation of PFC-amygdala connectivity in subjects with a history of MDD. The current study aimed to investigate the use of connectivity-based rtfMRI neurofeedback during an emotion regulation task in healthy subjects and subjects with a history of MDD. Thus, this study trained subjects to directly self-control the functional connectivity of two brain regions (one of the PFC regions and the left or right amygdala) with rtfMRI-nf in healthy controls and participants with a history of MDD. We hypothesized that there would be an overall increase in the functional connectivity between the amygdala and the PFC for both groups and that a greater increase would be seen in the HC group. We further hypothesized that the positive PANAS score would increase and the negative PANAS score would decrease after neurofeedback training for all participants, while the change would be more remarkable for the healthy controls. By training the functional connectivity that are selected individually through neurofeedback, we are hoping to find a more effective therapeutic emotion-regulation protocol for patients with the MDD.

## 2. Materials and Methods

### 2.1. Participants

Eleven young adult participants with a history of MDD (MDD group) and twelve healthy young adults matched for demographic variables (HC group) were recruited for this study. Subjects were excluded due to MRI incompatibility (e.g., metal implants, pregnancy, claustrophobia, etc.), history of neurological disorders, and brain trauma. In addition, subjects with a history of MDD were excluded if they suffered from an episode of depression in the month prior to the study date, and healthy subjects were excluded if they had family history of psychiatric disorders or any previous psychiatric diagnosis. All participants signed informed consent before taking part in the research and were compensated at the end of the study. The study received approval from the ethics committees of the College of Physicians & Surgeons, Columbia University, and the New York State Psychiatric Institute (NYSPI).

### 2.2. Psychological Assessments

Participants' MDD status was assessed at the time of the study by the depression module of the Young Adult version of the Diagnostic Interview Schedule for Children (DISC-YA; Shaffer et al., 2000), which assessed the past year and whole life separately; other psychiatric disorders were self-reported. Subjects' mood state was assessed by the Positive Affect Negative Affect Schedule (PANAS; Watson et al., 1988; Leue and Lange, 2011). In addition, the use of alcohol and/or drugs were assessed with the brief Michigan Alcohol Screening Test (MAST; Friedrich et al., 1978; Pokorny et al., 1972) and the Drug Abuse Screening Test (DAST-10; Skinner, 1982), respectively, to provide indexes for alcohol or drug abuse problems that could impact the behavioral and imaging data. Handedness laterality was assessed with the Edinburgh Handedness Inventory (Oldfield, 1971). Subjects were asked about their subjective feeling of emotional control over runs at the end of the study (results not provided here).

### 2.3. MRI Data Acquisition

**2.3.1. fMRI Data Acquisition**: Functional data were acquired on a GE Discovery MR750 3.0 Tesla scanner using a 32-channel coil, located in the MRI Unit, Department of Psychiatry, Columbia University-NYSPI. A gradient-echo T2*-weighted echo planar imaging (EPI) for blood oxygen level-dependent (BOLD) contrast pulse sequence was used for fMRI runs. Forty-five contiguous axial slices were acquired along the AC-PC plane, with an in-plane resolution of 3 ×3 mm and slice thickness of 3 mm (in-plane matrix size 64 × 64, TR = 2000 msec, TE = 25 msec, flip angle = 77°). Five dummy scans were used to allow the fMRI signal to reach a steady state.

**2.3.2. T1 Data Acquisition:** Structural data was acquired using a 3D T1-weighted fast spoiled gradient recalled (FSPGR) pulse sequence with isomorphic resolution (1 × 1 × 1 mm; 256 × 256 matrix, 180 slices, TI = 500ms, flip angle = 11°).

## 2.4. Functional Localizer

As detailed in previous studies (Bruhl et al., 2014; Linden et al., 2012), the emotional brain regions were functionally localized combined with anatomical toolboxes by presenting negative and neutral emotional stimuli from the International Affective Pictures System (IAPS; Lang, 2005) presented in 20 second blocks (**Figure 1**). The five neutral blocks and five negative blocks were pseudo randomized and alternated with the 10 fixation resting blocks consisting of 20 seconds of crosshair. Subjects were instructed to passively observe the pictures in the same order. Two localizer runs of 6 min 40 sec were acquired for each subject. This individual-based approach is designed to reduce the learning time for each subject when compared to targeting areas anatomically defined (Johnston et al., 2010).

## 2.5. Neurofeedback Task

Each subject participated in a single experimental session consisting of 4 runs of neurofeedback training (NF1 to NF4) and one transfer run without neurofeedback (**Figure 1**). Each neurofeedback run lasted for 5 min and 28 seconds and was consisted of a sequence of resting, emotional pictures, and counting backwards blocks; such a process repeated 5 times. During the resting block, subjects were instructed to "just look at the cross and relax" for the 12 seconds duration of the block. The initial baseline had an additional 28 seconds. The emotional blocks were consisted of 3 pictures of negative valence shown for 12 seconds each and a total of 36 seconds. Negative pictures with a negative valence mean of 2.66 and std 0.70 and arousal mean of 5.84 and std 0.74 were taken from the IAPS (natural disaster, animal, and human sets) and presented only a single time in the whole experimental session. The connectivity between amygdala and prefrontal cortex from the region of interest detected in the localizer run were calculated at real time and provided to the subjects for neurofeedback as a bar in the screen next to the stimuli. Subjects were instructed to control their emotions, which referred to decrease negative emotions, during the presentation of the emotional pictures using the movement of the neurofeedback bar and distancing strategies adapted from previous studies that used self-referencing strategies (Ochsner et al., 2004). Specifically, participants were told, *"You can take a few steps back mentally to move away from the picture and feel less involved with it. Or you can act like an objective observer without getting caught up in the emotions. Alternatively, you can eliminate any personal relationship to the picture."* These instructions were given outside the scanner and shown again on the screen prior to each run. During the counting backwards block subjects were told to mentally count backwards starting from the first presented number randomly selected from 100 to 200 (e.g., 102, 101, ...). During the 12 seconds, the counting cue was presented on the screen. The counting backwards blocks ensure that subjects disengage from the emotional control task (Zotev et al., 2011).

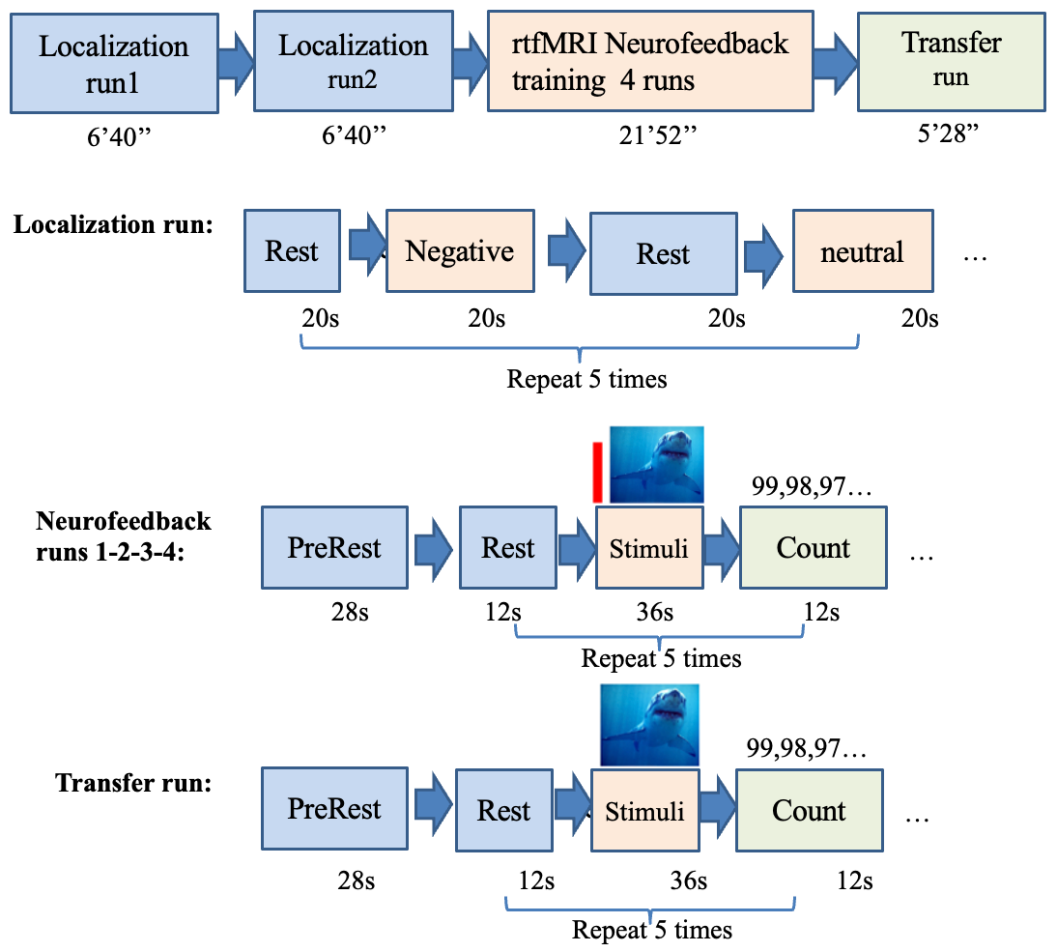

Figure 1. Experimental protocol for the Localization, Real-time fMRI Neurofeedback training and transfer runs

## 2.6. FMRI Analysis and Statistics

**2.6.1. Online fMRI Data Analysis:** All online data analysis was done using Analysis of Functional NeuroImages (AFNI; http://afni.nimh.nih.gov/; Cox and Hyde, 1997) and in-house software within the framework of the General Linear Model (GLM; Friston, 2005). **a) <u>Localization of the emotion regulation regions at the individual level:</u>** Preprocessing of each localizer run was done online immediately after acquisition and included correction for differences in slice timing, motion correction (i.e., realignment) to adjust for subjects' head movement, co-registration with the mean image of motion corrected EPI images to the individual FSPGR image (i.e., T1 image), and smoothing was performed by a Gaussian kernel 8 mm$^3$ FWHM (the full width at half maximum). A 128s temporal high-pass filter was applied to the preprocessed data to remove low-frequency noise for the functional localizer data. The fMRI data was analyzed in the local subject space in order to speed up. To correctly identify the anatomical location of BOLD signal, a MNI T1 image (template) was normalized to the subject's local space (i.e., register MNI T1 to subject's T1, resampled at 3 × 3 × 3 mm; local space resolution). First level data analysis was conducted after preprocessing using whole brain analysis and GLM model. Head movement parameters were used as regressors of no interest. Activation for the contrast between negative and neutral conditions within the amygdala and subdivisions of the frontal cortex selected based on previous findings (Young et al., 2014; Zotev et al., 2013) were used to determine the region of interest (ROIs) for the neurofeedback runs. For this purpose, masks of amygdala and preselected regions of the prefrontal cortex from the Automated Anatomical Labeling (AAL; Tzourio-Mazoyer et al., 2002) atlas were mapped to individual subject space. When multiple clusters in amygdala or regions of the prefrontal cortex were active, the cluster with larger percent of activation, as defined as cluster size divided by that AAL region size, were chosen as ROIs. Masks for the rt-fMRI-nf were created based on the activated cluster ($p < 0.05$, uncorrected) within that region. **b) <u>Real-time neurofeedback</u>**: Feedback from the emotion regulation task implemented using the in-house real-time fMRI system based on the real-time features of AFNI (Cox and Jesmanowicz, 1999) and a data transfer software (from the GE scanner to our real-time server) adapted from Dr. J. Bodurka (Laureate Institute for Brain Research) Real-time data analysis including motion detection and correction, and spatial smooth, which was conducted in native space. In our rtfMRI paradigm, the AFNI real-time plug-in was used to export mean values of fMRI signals for the ROIs in the emotional and prefrontal regions. Correlation of BOLD time series of between amygdala and frontal cortex areas were calculated in real time. The connectivity between amygdala and PFC regions identified during the first two runs of emotional task was the source of rtfMRI neurofeedback signal. Functional connectivity was calculated across those two activated brain regions, i.e., one picked up from the amygdala (either left or right hemisphere), another picked up from the frontal cortex area (either orbitofrontal or dorsal medial frontal, including anterior cingulate). Functional connectivity was defined as statistical dependencies between two regions by testing the Pearson correlation between two time courses (Banks et al., 2007; Friston, 2011; Horwitz et al., 1999; Stephan et al., 2008). The absolute value of the correlation coefficient was represented on the screen as a red bar, with a coefficient of 1 corresponding to the maximum of the bar height and 0 to no bar. The neurofeedback bar was updated every 2 seconds. In order to reduce the fluctuations of the feedback bar due to noise in the BOLD signal, the connectivity during the neurofeedback block was calculated based on a sliding time window of 20 time points, starting with the initial rest block at the beginning of the run.

**2.6.2. Off-line fMRI Data Analysis:** Localizer runs were analyzed off-line using SPM12 (http://store.elsevier.com/product.jsp?isbn=9780123725608). Preprocessing included slice timing correction, motion correction (i.e., realignment), co-registration of individual T1 image, normalization to MNI space and 8 $mm^3$ FWHM Gaussian kernel (the full width at half maximum) smoothing. A 128s temporal high-pass filter was applied to the preprocessed data to remove low-frequency noise for the functional localizer data. First level individual subject data analysis was conducted using whole brain analysis GLM model including head motion parameters as regressors of no-interest. The contrast between negative and neutral conditions within the amygdala and subdivisions of the frontal cortex was used to corroborate activation in amygdala and prefrontal ROIs selected during the NF and transfer runs.

**2.6.3. Statistical Analysis:** Before statistical analysis, outlier data points for BOLD activity in the ROIs greater than 3 standard deviations (SD) from the mean were excluded. The neurofeedback training effect was evaluated via a linear mixed-effects model with the fixed factors of neurofeedback run (NF 1 to 4), and group (MDD and HC) for amygdala-frontal connectivity changes using SAS Version 9.4 for Windows. Copyright © [2002-2012] SAS Institute Inc., Cary NC, USA, and adjusted for age. Run was entered as a linear term in order to test for a time trend, and a run × group interaction term was included to test whether the training effect differed by group. Associated t-statistics were used to assess the significance of main effects and interactions. Additionally, the model was re-fit using a categorical term for run. T-tests for the significance of differences between NF4 and NF1, NF4 and NF2, NF3 and NF2, and NF4 and the transfer run were performed based on the model in both MDD and HC subjects. Similar analyses were conducted to evaluate BOLD activity changes in amygdala and prefrontal ROIs. Associations of MDD status with PANAS scores (pre, post, and change) and MDD status, changes in connectivity, and their interaction with changes in PANAS scores were assessed using linear regression models. The associations between neurofeedback changes, change in emotion regulation, and change in PANAS score were determined via correlation analysis (Bernal-Rusiel et al., 2013a; Bernal-Rusiel et al., 2013b).

## 3. Results

### 3.1. Participants

One MDD participant was excluded from the study because the individual could not complete the scan protocol. Results are reported for 10 MDD participants and 12 HC participants. The mean age among the MDD group was 20.1 ± 1.1, and the mean age for the HC group was 19.2 ± 0.9 ($p = 0.04$). No significant differences in the distributions of sex, ethnicity, or handedness were observed between the two groups (**Table 1**). As expected, the MDD group had a significantly higher mean symptom count than the HC group. All participants scored below the threshold for alcohol or drug abuse problems in the DAST and MAST questionnaires.

|  | **MDD** (n = 10) | **HC** (n = 12) | **p-value*** |
|---|---|---|---|
| **Demographic Characteristic** |  |  |  |
| **Age**, years, mean (stdev) | 20.1 (1.1) | 19.2 (0.9) | 0.04 |
| **Sex** |  |  | 0.20 |
| Male, n (%) | 3 (30) | 8 (67) | |
| Female, n (%) | 7 (70) | 4 (33) | |
| **Ethnicity** | | | 0.77 |
| African American, not Hispanic, n (%) | 2 (20) | 1 (8) | |
| White, not Hispanic, n (%) | 0 (0) | 1 (8) | |
| Hispanic, n (%) | 8 (80) | 10 (83) | |
| **Neuropsychological Characteristics** | | | |
| Handedness, mean (stdev) | 84.7 (17.6) | 94.1 (11.1) | 0.14 |
| DISC-YA MDD symptom count, mean (stdev) | 14.7 (6.3) | 6.1 (4.4) | 0.001 |

Abbreviations: DISC-YA: Young Adult Diagnostic Interview Schedule for Children;

*p-value from 2 sample (MDD, HC) t-test or Fisher's Exact Test.

Table 1. Demographic, and Neuropsychological Characteristics of Participants

### 3.2. ROI Localization

Passive viewing of negative compared to neutral images during the emotional task activated amygdala in 68% of participants (**Figure 2**). Based on our ROI criteria selection using the ratio of cluster size to each anatomical mask, we chose the right amygdala in most cases. Additionally, if activation was not detected, an anatomical-based ROI in the right Amygdala would be chosen. Activations for the frontal region were observed in several anatomical regions of the frontal cortex and anterior cingulate during the localizer runs as observed in other studies (Bruhl et al., 2014). Based on our ROI selection criteria, we selected ROIs in the superior frontal areas (orbitofrontal and medial), medial orbitofrontal, or anterior cingulate cortex for each subject. Based on previous research work (Ochsner et al., 2004), although the functions of the prefrontal cortex are diverse, mPFC is for conceptual evaluation, lPFC, cingulate PFC, vlPFC and dorsolateral PFC regions are all involved in cognitive control, which all participated in the emotional regulation. Anterior cingulate is also supported to be participating in cognitive emotional regulation (Giuliani et al., 2011).

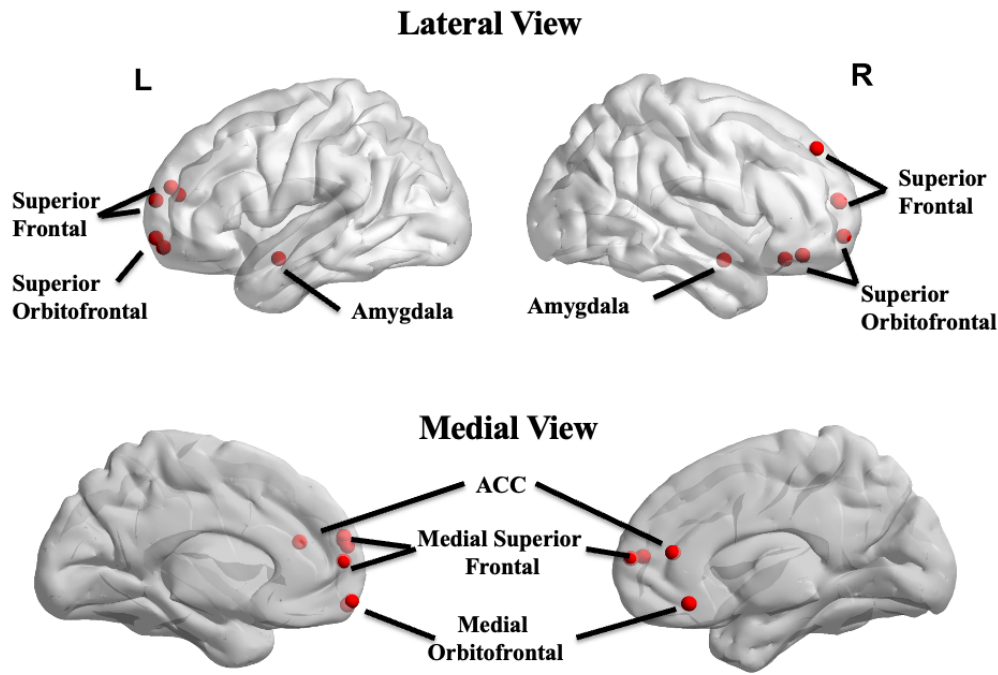

Figure 2. Localization results. Individual region of interest (ROI) for amygdala and prefrontal cortex for all subjects which were transferred and overplayed on a single brain of MNI space for visualization purpose. ACC: anterior Cingulate Gyrus

**3.3. Functional Connectivity During Emotion Regulation Blocks in Neurofeedback and Transfer Runs**

Functional connectivity changes across four rtfMRI-nf runs showed a marginally significant interaction between the MDD history +/- status and training time (runs; $p = 0.097$). MDD participants, but not healthy controls, show a trend of decreasing functional connectivity after neurofeedback trainings ($p = 0.085$; **Figure 3**). Connectivity changes between NF4 and NF1 showed a close to significant difference ($p = 0.06$) in the MDD group only. Functional connectivity changes across the last rtfMRI-nf run and the transfer run showed no significant difference either in the MDD ($p = 0.69$) or the HC ($p = 0.86$) group. Although functional connectivity in MDD group was higher than in healthy controls, no significant connectivity strength difference was observed between MDD and HC groups at NF1. There were no significant functional connectivity changes between amygdala and prefrontal cortex during counting backwards as expected.

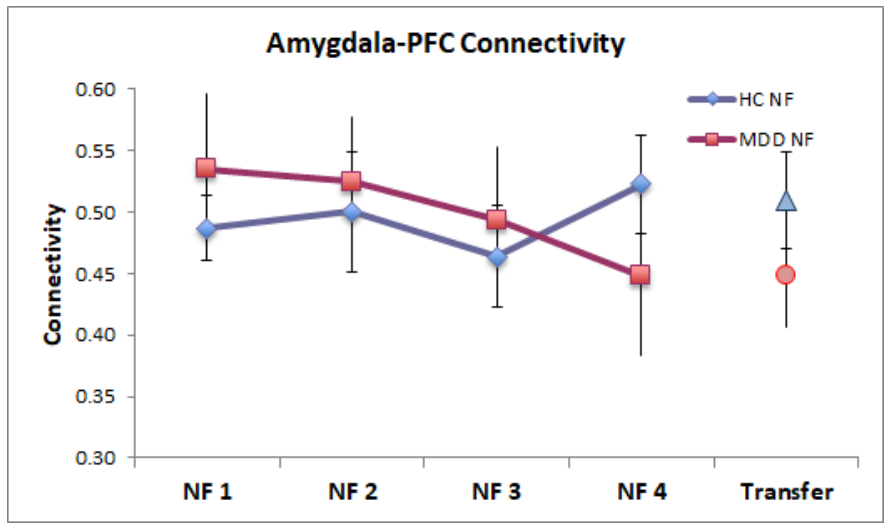

Figure 3. Linear Mixed Effects Model for Connectivity between amygdala and PFC area. HC= blue, MDD=red, 1-4 are rt-fMRI-NF runs, 5th is the transform run without feedback.

*Connectivity refers to the correlation coefficients from Pearson correlation

### 3.4. BOLD activation during emotion regulation blocks in Neurofeedback and Transfer runs

The linear mixed-effects model revealed no significant differences in BOLD activation neither in amygdala nor prefrontal cortex for NF1-NF4 in MDD or HC groups or a run × group interaction (**Supplementary Figures S1&S2**). T-tests between BOLD activity for NF2 and NF3 reveal an approaching significant difference for Amygdala ($p = 0.065$) in the MDD group and significant difference for Prefrontal Cortex ($p = 0.041$) in the HC group. There were no differences for amygdala or prefrontal cortex BOLD activity between NF4 and NF1, NF4 and NF2, NF4 and the transfer run in either group. No significant BOLD activation difference was observed between the MDD and the HC groups at NF1.

### 3.5. Psychological Assessments

The comparison of participants' emotional state, as measured by the PANAS prior to and following the rtfMRI-nf showed a significant decrease in negative PANAS scores for the whole sample (**Table 2**). In addition, positive PANAS scores decreased significantly in the MDD group. The MDD group showed larger pre-post decreases than did the controls for both positive and negative PANAS scores (**Table 2**). Nevertheless, these differences were not statistically significant between the MDD and the control group (**Table S1**). Linear regression models confirmed that history of MDD was associated with a higher negative PANAS score before ($p$

= 0.01) as well as after (*p* = 0.02) rtfMRI-nf and that adjusting models for pre-PANAS score did not alter the conclusions (**Supplementary Table S1**).

3.6. Relationship between Psychological Assessments and Connectivity Changes

Further analysis indicated that neither history of MDD nor amygdala-prefrontal changes in connectivity predicted the difference between pre and post PANAS scores, and that the effect of change in connectivity did not differ between the MDD group and the HC group (**Supplementary Tables S2 & S3**).

|  | Positive PANAS | | | | | Negative PANAS | | | | |
|---|---|---|---|---|---|---|---|---|---|---|
|  | **Pre** | **Post** | **Post-Pre Difference** | | | **Pre** | **Post** | **Post-Pre Difference** | | |
|  | *Mean (SD)* | *Mean (SD)* | *Mean (SD)* | *t* | p | *Mean (SD)* | *Mean (SD)* | *Mean (SD)* | *t* | p |
| **MDD** | 34.6 (5.6) | 31.1 (6.3) | -3.50 (3.69) | -3.00 | **0.02** | 16.6 (5.9) | 14.4 (5.0) | -2.20 (3.65) | -1.91 | 0.09 |
| **Controls** | 29.4 (8.1) | 28.2 (10.3) | -1.25 (5.33) | -0.81 | 0.43 | 13.0 (3.3) | 11.4 (2.8) | -1.58 (4.03) | -1.36 | 0.20 |
| **Full Sample** | 31.8 (7.4) | 29.5 (8.7) | -2.27 (4.69) | -2.27 | **0.03** | 14.6 (4.9) | 12.8 (4.1) | -1.86 (3.78) | -2.31 | **0.03** |

Table 2. Mean PANAS scores by group and overall, and Paired T-tests for Post- and Pre-rtfMRI-nf training differences.

4. Discussion

The present study used an individual-based approach to evaluate the feasibility of employing connectivity between amygdala and prefrontal cortex as a concurrent neurofeedback signal to regulate emotions in healthy controls and subjects with a history of MDD. The connectivity-based neurofeedback training resulted in a trend of decreasing functional connectivity between amygdala and prefrontal cortex in subjects with a history of MDD over the four rtfMRI neurofeedback training runs. Moreover, we also observed a decrease of positive PANAS score in the MDD group compared to the PANAS score before rtfMRI-nf training.

Current theories of emotion regulation propose two fundamental processes during processing of emotional stimuli, top-down cognitive regulation and bottom-up encoding of the affective stimuli (Ochsner et al., 2009; Urry et al., 2006). In healthy individuals, adaptive emotion regulation in the presence of negative valence emotional stimuli engages an inhibitory action of the prefrontal cortex over the amygdala (Johnstone et al., 2007; Urry et al., 2006; Vanderhasselt et al., 2013). The increased activation in prefrontal cortex and decreased activation in amygdala was also observed during conscious intentional emotion regulation strategies, like adaptive reappraisal of negative emotion, in healthy subjects (Ochsner et al., 2004; Phan et al., 2005). Structural MRI studies have shown direct connection between amygdala and vmPFC and lateral/dorsal PFC (Price, 2005), suggesting that PFC-amygdala circuitry is a hub-relay for emotion regulation. A few previous studies on rtfMRI neurofeedback aimed at emotion regulation have been successful in regulating amygdala activity (Herwig et al., 2019; Posse et al., 2003; Young et al., 2018; Yuan et al., 2014; Zotev et al., 2011; Zotev et al., 2013; Zweerings et al., 2020). Importantly, off-line post hoc analysis of the fMRI data in rtfMRI-nf have shown that self-regulation of a single ROI is accompanied by modulations in the connectivity of whole brain networks (Hamilton et al., 2011; Koush et al., 2013; Lee et al., 2012; Lee et al., 2011; Rota et al., 2011; Ruiz et al., 2014; Ruiz et al., 2013; Scharnowski et al., 2014; Veit et al., 2012; Zilverstand et al., 2014; Zotev et al., 2011). For example, increased prefrontal-AMG connectivity was observed after post hoc analysis of neurofeedback training (Zotev et al., 2013). In this study, the negative emotional pictures presented during the neurofeedback training were expected to activate amygdala and engage inhibitory frontal control over amygdala in healthy subjects, subsequently increasing functional connectivity between these regions. Rt-fMRI-connectivity neurofeedback was expected to help increase volitional control over negative emotions via enhanced connectivity between these regions. Although we did not observe the expected effect of neurofeedback on connectivity modulation in our healthy control group, there was a significant increase in prefrontal cortex between the second and third neurofeedback runs. Such a change suggested that healthy controls attempted to increase frontal activation to regulate their emotion over the NF runs.

On the other hand, MDD is characterized by emotion dysregulation associated with attenuated cognitive control in PFC, high amygdala response to negative stimuli (Johnstone et al., 2007; Sheline et al., 2001), and decreased functional connectivity (task-related and resting-state) as well as structural connectivity in multiple networks including emotion regulation network (Anand et al., 2009; Carballedo et al., 2011; Connolly et al., 2013; Frodl et al., 2010; Geng et al., 2016; Greicius et al., 2007; Heller, 2016; Kaiser et al., 2015; Klauser et al., 2015; Kong et al., 2013; Liao et al., 2013; Lui et al., 2011; Murrough et al., 2016; Ramasubbu et al., 2014; Sambataro et al., 2014; Sheline et al., 2010; Tang et al., 2013; Zhu et al., 2012). A previous study also revealed decreased functional connectivity between right amygdala and right ventrolateral frontal cortex in female MDD patients compared to controls (Yang et al., 2017). Similarly, among elderly patient with late onset depression, decreased resting state functional connectivity from left amygdala to right middle frontal gyrus and the left superior frontal gyrus was observed (Yue et al., 2013). However, although some studies on individuals with remitted depression show patterns of increased amygdala reactivity, decreased prefrontal activity and dysregulation (Dore et al., 2018; Kanske et al., 2011), other studies suggest an increased amygdala-PFC connectivity. Indeed, functional connectivity was shown to be increased in elderly females with remitted MDD in response to negative stimuli; an interpretation could be related to increased attention bias to negative images in these subjects (Albert et al., 2017). Furthermore, adolescents with a history of MDD also showed hyperconnectivity from the left amygdala to the right PCC, which was positively correlated with higher rumination (Peters et al., 2016). The difference in amygdala-PFC connectivity in subjects with a history of MDD among studies may be due to the differences in depression severity or the early/later onset of MDD in those samples. In this study, we observed similar connectivity between PFC and amygdala in subjects with a history of MDD compared to HC at the first Neurofeedback run.

However, as the task progressed, the MDD group showed a trend in decreasing PFC-amygdala connectivity with no significant changes in either frontal or amygdala activity, indicating that they failed to engage the frontal cortex consistently, which resulted in the observed decreased connectivity over the NF runs. These results suggested that subjects with a history of MDD failed to use the connectivity feedback. Indeed, the concurrent feedback may have elicited negative emotions (insecurity, fear to fail, etc.) rather than providing a positive regulation feedback according to their negative cognitive biases to interpret stimuli (Beck, 2008). In addition, there was no significant difference between amygdala or prefrontal activation in NF1 between groups.

Our participants were not currently in a depression episode, and their level of amygdala reactivity and prefrontal activation, indicated by BOLD at NF1, seemed fairly similar to that of healthy controls.

Moreover, diminished positive affect and sustained negative affect are central characteristics of MDD. These MDD symptoms have been proposed to be the result of difficulties in emotion regulation (for a review, see Joormann and Quinn, 2014). In this study, we provided subjects with connectivity-based neurofeedback as an index of their emotion regulation network status while subjects observed negative images. We compared participants' emotional state, as measured by the PANAS, before and after the rtfMRI-nf as a measure of emotional regulation success. Importantly, different cognitive strategies have been shown to modulate negative and positive affect in varied ways in everyday life. It has been revealed that distancing strategy increases positive affect and decreases negative affect, but less than other strategies (Rood et al., 2012). Other scholars observed that rumination and suppression were related to decreases in positive affect and increases in negative affect, reflection was associated with an increase in positive affect only, and reappraisal, distraction, and sharing were associated with increase in positive affect (Brans et al., 2013). In addition, several studies have documented decrease in negative affect after reappraisal (Smoski et al., 2013; van der Meer et al., 2014). Since subjects were instructed to distance themselves from the negative stimuli, we expected both groups to increase engagement of emotional regulation mechanisms over time to diminish their negative emotions and possibly increase their positive emotions. A related decrease in negative PANAS score and increase in positive PANAS score were also expected. The results showed that there was a significant decrease in negative PANAS scores in the whole sample. The decrease in negative PANAS score overall suggests that the rt-fMRI-nf training helps with modulating negative affect. However, this decrease was not coupled with an increase in connectivity between amygdala and prefrontal cortex. There was also a significant positive PANAS decrease among the whole sample that may reflect the significant decrease in positive PANAS scores for the MDD group after the four rt-fMRI-nf runs. Although a decrease in positive PANAS score in the MDD group was not expected, it followed the trend of decreasing connectivity between amygdala and prefrontal cortex observed in these subjects over runs. Overall, it suggests that participants with a history of MDD were unable to engage the emotion regulation mechanism during the task. As previously suggested, the variability of the connectivity signal used as concurrent feedback during the negative images may have been associated with negative emotions (increased insecurity, fear to fail, etc.) in the MDD group leading to a decrease in positive affect. This is in agreement with the observation that depressed subjects have difficulties to apply self-distancing from negative events and reappraisal as compared to healthy individuals (Gunaydin et al., 2016).

## 5. Limitations

Some general issues concerning the methodology of the current study need to be addressed. First, the sample size was relatively small to generalize results to a population level and could result in reduced the statistical power. The results should be further validated in a larger randomized-control trial study. In addition, MDD and HC groups were not matched by sex, which may limit the interpretation of the statistical comparison. Second, the number of neurofeedback training sessions was limited to one session with four runs, only allowing to test the feasibility of the technique and observe short-term changes. However, the data suggested that a single session was not enough to benefit from the neurofeedback training. Remitted MDD participants may require more sessions of the rtfMRI-nf training in order to be accustomed to being exposed to negative stimuli and enhance the control of the emotion network when confronted with negative stimuli. Multiple sessions should be arranged to investigate the accumulative effect of neurofeedback training on brain connectivity and PANAS scores. The future study would also benefit from precisely deriving the direction of change in PFC-amygdala, since we did not precisely give the subjects direction of change to avoid possibly aggravating our subjects' condition after the neurofeedback training. Third, the study lacks a sham group, such as selecting other brain regions (e.g., parietal cortex) as the target for rtf-MRI-nf training, which could be helpful for determining whether stimulating other regions can yield similar effect on the emotion processing. However, neurofeedback from areas not associated with the task at hand can also confuse the subject and hinder task performance.

Finally, two essential modifications from previous connectivity-based studies were implemented here that may also account for the observed results. First, unlike most published connectivity-based studies (Koush et al., 2013; Megumi et al., 2015; Zilverstand et al., 2014) that adopted the neurofeedback signal immediately after the training epoch, this study provided concurrent neurofeedback as a red bar that was updated every 2 seconds, similar to that in Ramot et al.'s study (2017). This approach was based on numerous ROI-based rtfMRI studies (for a review, see Ruiz et al., 2014) that indicated that neurofeedback presented during the regulation epoch led to a strong training effect. Second, we chose to select the prefrontal ROI based on subjects' individual response to the emotional localizer blocks as in Linden and colleagues' study (2012). The reasons for this approach were two folds. First, studies of emotion regulation in HC subjects suggest strong interactions between a number of prefrontal regions and amygdala, including ventromedial PFC, dorsomedial PFC, dorsolateral PFC, orbitofrontal and ACC (Banks et al., 2007; Hallam et al., 2015; Hariri et al., 2003; Kanske et al., 2011; Ochsner et al., 2004; Stein et al., 2007; Zotev et al., 2013). Furthermore, changes in connectivity between amygdala and those areas have been observed after emotion regulation neurofeedback sessions (Ruiz et al., 2013; Zotev et al., 2011). Second, we focused on the feasibility of concurrent connectivity-based NF training rather than on participants' ability to learn to regulate a path between specific brain regions. Indeed, former studies suggest that different pathways may be invoked according to the regulation strategy used by the participant (Kanske et al., 2012; Ochsner et al., 2004). Therefore, the particular PFC ROI for each subject was based on the strongest activation cluster during the localizer run to maximize the connectivity signal. The obtained ROIs coincide with current meta-analysis of amygdala-PFC connectivity (Robinson et al., 2010). Although the use of individualized ROIs was considered an advantage to target subject specific pathways, it also leads to variability in the NF efficacy as some pathways may be involved in upregulating amygdala. Indeed, Ochsner et al. (2004) reported the involvement of rostromedial PFC for the retrieval of emotion knowledge while lateral and orbital PFC were involved in amygdala down regulation. Nevertheless, though Ochsner and Gross (2014) state that the functions of the prefrontal cortex are diverse, mPFC is for conceptual evaluation, lPFC, cingulate PFC, vlPFC and dorsolateral PFC regions are all involved in cognitive control, collectively participating in the emotional regulation. In addition, the study differentiated the functions of different PFC areas based on the strategy used and specifically reported an increase in most PFC areas for reappraisal (Ochsner and Gross, 2014). In our study, we used a version of self-referencing strategy called distancing strategy, which is a type of reappraisal strategy (Ochsner et al., 2004).

Another limitation regarding the choice of ROI is that we used a passive localization run to identify ROIs to use for emotion regulation training. During the localization run, only the participants' PFC regions were activated, and no amygdala activity was shown during emotion regulation. Instead, we tried to identify individual-based emotion processing areas through passive localization run.

## 6. Conclusions

This study extends the scope of rtfMRI-nf research from a single ROI to functional connectivity of emotional networks using concurrent feedback signals and individual-based ROIs. The trend of decreased amygdala-PFC suggests that concurrent connectivity signals during the regulation task may be distracting for emotion regulation in subjects with a history of MDD. These findings are in agreement with previous studies concluding that youth with early onset depression showed reduced capacity for feedback effects on ROI's and their connectivity. On the other hand, the increase of prefrontal activity between neurofeedback runs 2 and 3 suggests that this technique may help healthy subjects to recruit the frontal cortex for volitional emotion regulation. Further studies will be necessary to determine the effectiveness of connectivity-based rtfMRI feedback in subjects with a history of MDD in a large sample with more sessions of rtfMRI-nf training.

**Acknowledgments**

This study was supported in part by a NARSAD Young Investigator Grant (PI: X. He) from the Brain & Behavior Research Foundation, The New York State Psychiatric Institute MRI Pilot Award (PI: X. He), Columbia Irving Institute Imaging Pilot Award (PI: X. He) from NIH UL1 TR000040, and a NIDA R01 (1 R01 DA038154-01A1, PI: C. Hoven). We also thank Drs. Jeff Miller, Lupo Geronazzo, and Bryan Denny for the emotional task and the cognitive strategy.

Supplementary Materials:

**Table S1. Linear regression models: MDD status predicting pre, post, as well as the post and pre-differences in PANAS scores (n=22).**

| | Positive PANAS | | | Negative PANAS | | |
|---|---|---|---|---|---|---|
| **MDD status predicting** | *Beta* | *SE* | *p* | *Beta* | *SE* | *p* |
| **Pre PANAS** [a] | 3.89 | 3.40 | 0.27 | 5.53 | 2.03 | **0.01** |
| **Post PANAS** [a] | 2.05 | 4.24 | 0.64 | 4.54 | 1.73 | **0.02** |
| **Post-Pre difference** [a] | -1.84 | 2.26 | 0.43 | -1.00 | 1.87 | 0.60 |
| **Post-Pre difference** [b] | -2.06 | 2.39 | 0.40 | 2.11 | 1.80 | 0.26 |

Abbreviations: PANAS: Positive Affect Negative Affect Schedule. [a] Models adjusted for age. [b] Models adjusted for age and the pre- PANAS score

**Table S2. MDD status, Pre-Post Connectivity Changes, and their interaction predicting pre-post differences in Positive PANAS scores (n=22).**

| Parameter | Estimate | Standard Error | t Value | Pr > \|t\| |
|---|---|---|---|---|
| **Intercept** | -2.66 | 5.16 | -0.52 | 0.61 |
| **MDD** | -2.64 | 2.58 | -1.02 | 0.32 |
| **Post-Pre Connectivity** | -2.98 | 13.35 | -0.22 | 0.83 |
| **Post-Pre Connectivity *MDD** | -5.69 | 16.89 | -0.34 | 0.74 |
| **Pre- PANAS positive score** | 0.05 | 0.17 | 0.33 | 0.75 |
| **Age (centered)** | -0.74 | 1.17 | -0.63 | 0.54 |

**Table S3. MDD status, Post-Pre Difference in Connectivity, and their interaction predicting pre-post differences in Negative PANAS scores (n=22).**

| Parameter | Estimate | Standard Error | t Value | Pr > |t| |
|---|---|---|---|---|
| Intercept | 6.90 | 2.49 | 2.77 | 0.01 |
| MDD | 2.06 | 1.74 | 1.19 | 0.25 |
| Post-Pre Connectivity | -14.82 | 7.99 | -1.85 | 0.08 |
| Post-Pre Connectivity*MDD | 12.41 | 10.07 | 1.23 | 0.24 |
| Pre- PANAS negative score | -0.61 | 0.17 | -3.58 | 0.003 |
| Age (centered) | -1.08 | 0.81 | -1.34 | 0.20 |

**Figure S1.** Linear Mixed Effects Model on amygdala BOLD. HC= blue, MDD=red, 1-4 are rt-fMRI-NF runs, 5th is the transform run without feedback.

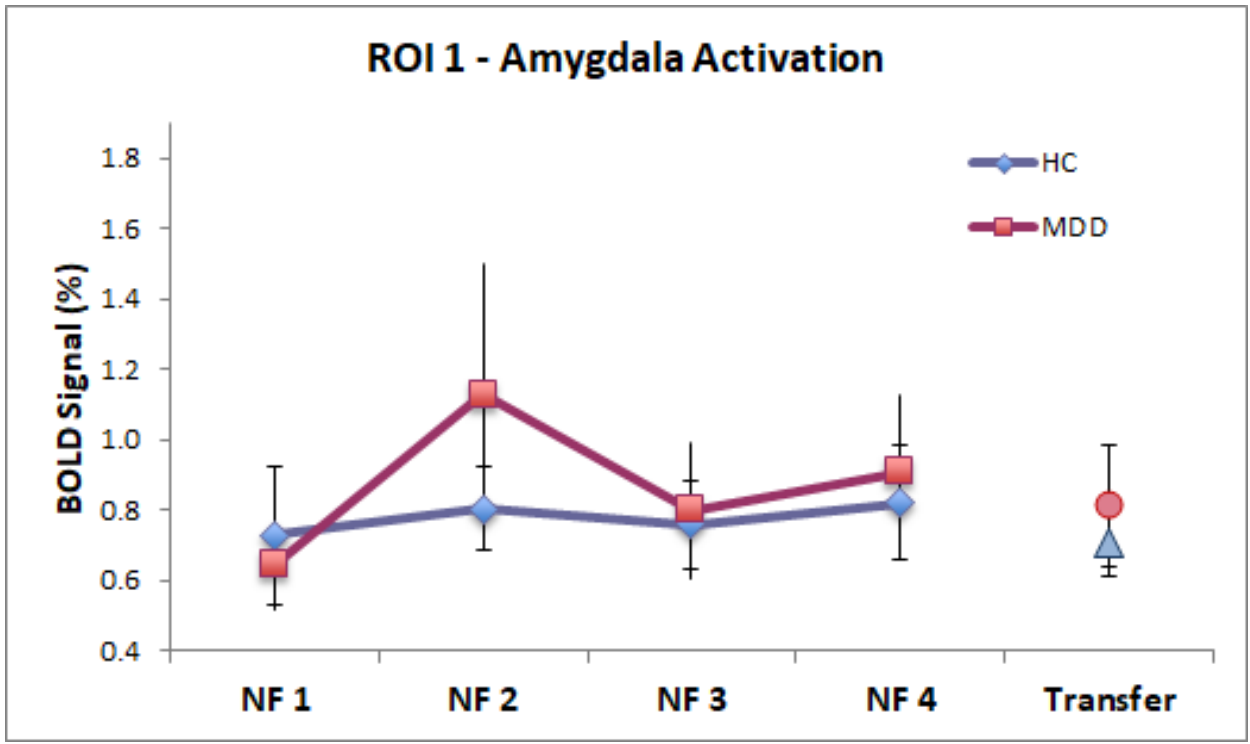

**Figure S2.** Linear Mixed Effects Model on PFC BOLD. HC= blue, MDD=red, 1-4 are rt-fMRI-NF runs, 5th is the transform run without feedback.

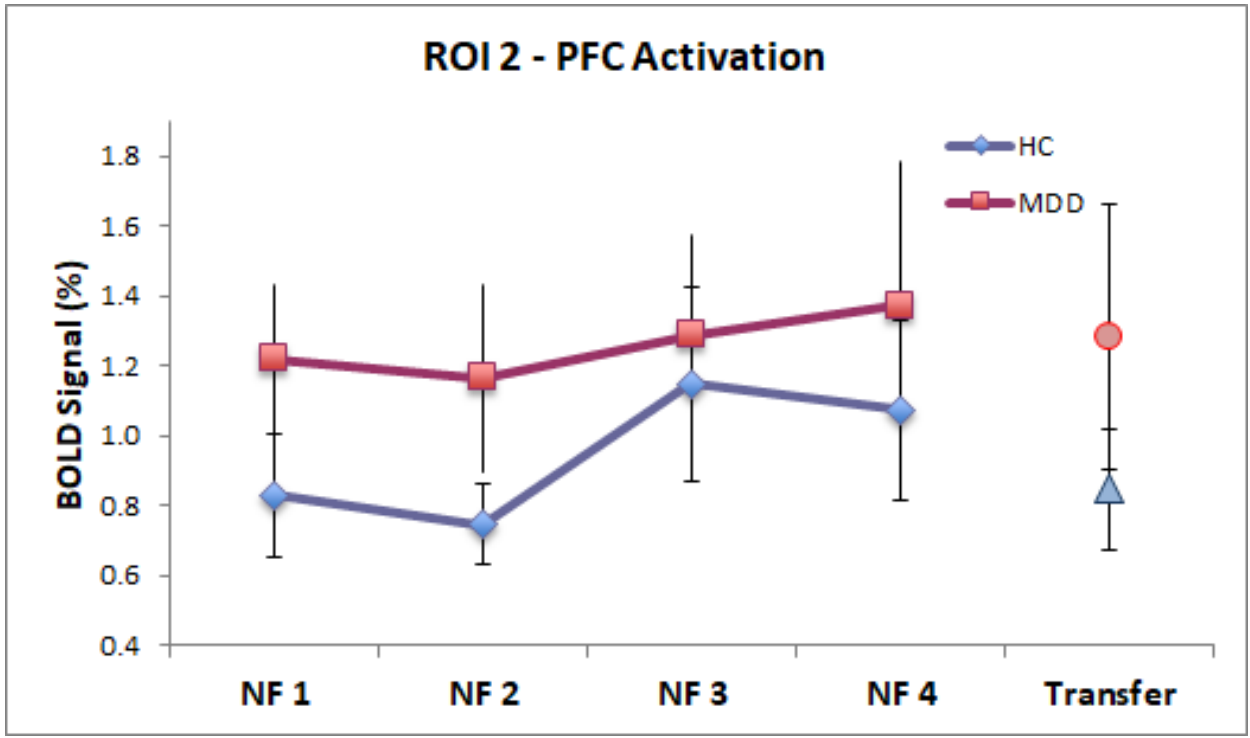